\documentclass[prl,twocolumn]{revtex4}
\usepackage{latexsym}
\usepackage{graphics}
\usepackage{epsfig}
\usepackage{amssymb}
\begin{document}
\title{Divide and conquer: resonance induced by competitive interactions}
\author{T. Vaz Martins$^{1,2}$, Ra\'ul Toral$^{2}$ and M. A. Santos$^{1}$}
\affiliation{(1) Departamento de F\'isica and CFP, Faculdade de Ci\^encias da Universidade do Porto, Rua do Campo Alegre 687,
4169--007 Porto, Portugal \\(2) IFISC (Instituto de F{\'\i}sica Interdisciplinar y Sistemas Complejos),
CSIC-UIB, Ed. Mateu Orfila, Campus UIB, E-07122 Palma de Mallorca, Spain}
\date{\today}
\begin{abstract}
We study an Ising model in a network with disorder induced by the presence of both attractive and repulsive links. This system is subjected to a subthreshold signal, and the goal is to see how the response is enhanced for a given fraction of repulsive links. This can model a network of spin-like neurons with excitatory and inhibitory couplings.
By means of numerical simulations and analytical calculations we find that there is an optimal probability, such that the coherent response is maximal. 
\end{abstract}
\maketitle
\section{Introduction}
\label{intro}
In nonlinear systems, the right amount of noise can amplify the response to a weak periodic signal. This constructive effect of noise, known as stochastic resonance, was first proposed to explain the occurrence of ice ages \cite{benzi,nicolis}, and has since found applications in such diverse areas as lasers \cite{lasers}, SQUIDS \cite{SQUIDS}, or neurons \cite{neurons}, just to mention a few \cite{GHJM98,JSTAT}. 

The mechanism of stochastic resonance involves a matching between the frequency of the external signal and a stochastic frequency induced by the noise. The prototypical system is one in a bistable potential subjected to a periodic modulation signal and to noise. A weak, subthreshold, signal cannot, by itself, make the system switch between its equilibrium states. When driven only by noise, the system is able to jump between the two minima, with a mean frequency which depends on the intensity of the noise as given by the celebrated Kramers' formula \cite{kramers}.  It is then possible to tune the noise intensity in order to match the noise induced jumping frequency with the frequency of the forcing. At this point, we have an amplified coherent response. Based on those simplified ingredients, stochastic resonance has been applied to many areas and situations, and extended from systems with a few degrees of freedom to systems composed by many units. 

It was shown in reference \cite{TMTG06} that diversity or heterogeneity, in the form of quenched disorder, can play the same constructive role as noise. The authors considered a system composed by many coupled bistable units, subjected to an external weak periodic signal. Diversity is introduced as variability of a parameter that controls the relative stability of each bistable state of the potential. When the units are identical (and both states are equally stable for all units) the signal is subthreshold and, because of the coupling, all units remain in the same state. As diversity increases, the signal becomes, for half of its period, supra-threshold  for some of the units and forces those units to jump from their less stable state to the other. In the other half of the period, the signal becomes supra-threshold for a different set of units. The units which follow the signal pull the other units, to whom they are attractively coupled, and the collective effect is that a significant fraction of the units is able to  respond to the external forcing.  For this mechanism to be effective, the units cannot be too diverse, because then some of them would offer too much resistance to follow the signal. But they can not be too  similar either, because then there wouldn't a big enough fraction that can follow the signal. The optimal value of diversity is the one that makes the system more sensitive to the external signal.  The collective effect can be understood as the result of the degradation of entrainment induced by diversity, a similar effect to that induced by noise\cite{TTV:2007}. This degradation results in the lowering of the effective potential barrier separating the two stable states of the collective system. The barrier can then be more easily overcome by the external forcing. One of the main points made in \cite{TMTG06} is that it is the loss of entrainment that drives the resonance effect. This loss of entrainment can be induced by diversity \cite{TMTG06}, by noise (in the case of extended stochastic resonance \cite{array,Wio}), or by some other source. Along these lines,  the role of the heterogeneous complex network topology in the amplification of external signals has been addressed in \cite{ALA:2007}, and Chen et al. \cite{CZL:2007} have shown how structural diversity enhances the cellular ability to detect extracellular weak signals. The interplay between noise and diversity in an ensemble of coupled bistable FitzHugh-Nagumo elements subject to weak signal has been considered in \cite{GGK:2007}. 

The presence of both repulsive and attractive interactions is not unusual in systems with many units. The existence of inhibitory and excitatory connections in the brain neurons, or a society with friends and enemies are examples of such systems. The emergence of a coherent behavior in the absence of forcing and in the presence of repulsive links was treated in \cite{Levya}. There it was shown that one can obtain a more coherent behavior, in the form of synchronized pulsing,  by adding an optimal amount of long-range repulsive couplings in a mixture of excitable and oscillatory units described by the Hodgkin-Huxley model. In the same reference, a similar improvement of the internal coherence in an Ising model with a simple majority-like dynamics in the presence of long-range repulsive links was also shown. Also in \cite{TZT08}, an intermediate amount of repulsive links was found to trigger collective firing in an ensemble of active-rotators \cite{shinomoto} in the excitable regime. The role of diversity in heterogeneous excitable media was considered in \cite{julyan} where the author demonstrates that diversity in a parameter can cause the emergence of global oscillations from individually quiescent elements in a system of van der Pol-FitzHugh-Nagumo elements. The combined effects of noise and variability in the synchronization of neural elements has been studied in \cite{GGK:2008}, while reference \cite{TSTC07} unveils the general mechanism for collective synchronized firing in excitable systems arising from degradation of entrainment originated either by noise, diversity or other causes.

In this work we study a periodically forced system where the only source of disorder is competitive interactions and show that competition in the sign of interactions may also lead to a resonance effect. This resonance can be interpreted as an optimal transmission of the information carried by the external signal, in a kind of ``divide and conquer'' effect.
The paper is organized as follows: In section \ref{sec:2} we introduce the model and present the results of numerical simulations that show the existence of the resonance effect; in section \ref{sec:3} a mean-field approximation, which is able to reproduce qualitatively the  simulations results is detailed; in section \ref{sec:4} we discuss in detail the mechanisms that may lead to the resonance, both from microscopic and macroscopic points of view; finally, in section, \ref{sec:5} we end with some brief conclusions and outlooks. 

\section{Model and results}
\label{sec:2}

\subsection{Model}
\label{sec:2.1}
We consider a set of $N$ spin-like (Ising) dynamical variables $\mu_i(t)$ which, at a given time $t$, can adopt one of two possible
values, $\mu_{i}=\pm1$. We will sometimes use the language of a magnetic system, but our aim is quite general and these states can represent, for instance, two different opinions (in favor/against) about a topic, the state of a neuron (firing/not firing), or several other interpretations  \cite{Levya,KZ02}. The variables are located on the nodes of a given network whose links represent interactions. We assign a weight $\omega_{ij}$ to the link connecting nodes $i$ and $j$ and consider only the symmetric case $\omega_{ij}=\omega_{ji}$ (or an undirected network). According to the discussion above, we let the weights take positive or negative values: $\omega_{ij}=1$ or $\omega_{ij}=-\kappa$ with $\kappa>0$. The neighborhood of node $i$ is the set $V(i)$ of nodes $j$ for which a connecting link between nodes $i$ and $j$ exists.

The spin variables evolve according to the following dynamical rule: At time $t$ one of the variables, say $\mu_i$, is chosen at random. The value of this variable is updated according to:
\begin{equation}
\label{par}
\mu_i(t+\tau)=
\left\{ \begin{array}{ll}
{\rm sign}\left[\sum_{j\in V(i)}\omega_{ij}\mu_{j}(t)\right] & \textrm{w.p.} \ 1-|a\sin(\Omega t)|,\\ \\
{\rm sign}\left[\sin(\Omega t))\right] & \textrm{w.p.} \  |a\sin(\Omega t)|,
\end{array}\right.
\end{equation}
(w.p. stands for ``with probability").
In both cases, if the expression within square brackets is equal to zero, the variable does not change: $\mu_i(t+\tau)=\mu_i(t)$. The first case represents a weighted ``majority-rule''  in which the state of the spin is determined by the sign of its {\it local field} $h_i(t)=\sum_{j\in V(i)}\omega_{ij}\mu_{j}(t)$. The second case represents the effect of an external forcing of frequency $\Omega$ -- the intensity $a<1$ determines the rate at which the signal influences the dynamics of the variable $\mu_i$. The choice of the time step $\tau=1/N$ defines the unit of time as $N$ updates. 
We consider both regular lattices (with $k$ neighbors) and random networks of the small-world type. The latter are constructed in the usual way \cite{watts}. Denoting by $q$ the rewiring probability (percentage of short-cuts), the limit $q=1$ corresponds to a random Erd\"os/Renye-type network, $q=0$ is a regular ring-network and intermediate values of $q$ define a small-world network. We have also considered a square lattice in which a node is linked to the $k=8$ nodes of its Moore neighborhood. In each case, links are assigned a strength $-\kappa$ with probability $p$ or a strength $1$ with probability $1-p$. In the case of a random network, the number of links (degree) $k_i$ of node $i$ is a random variable with probability $P_{k_i}$ and average $\langle k_i \rangle =k$. Denoting by $k_i^+$ and $k_i^-$  respectively the number of positive and negative links of node $i$, its degree is $k_i=k_i^+ + k_i^-$ and $\langle k_i^+\rangle=(1-p)k$, $\langle k_i^-\rangle=pk$.

It is worth noticing that, from the formal point of view, the majority-rule is equivalent to a heat-bath stochastic dynamics in the limit of zero temperature\cite{LB:2000}. The Hamiltonian is ${\cal H}=-\sum_{\langle i,j\rangle}\omega_{ij}\mu_i\mu_j$ (the sum runs over all pairs of neighbors) and the majority-rule always leads to a configuration with less or equal energy. If all the weights $\omega_{ij}$ are positive, the ground states are $\mu_i=+1$ or $\mu_i=-1$, $\forall i$, and these ground states are reached independently of the initial condition. If there is a fraction of negative links, the system is of the spin-glass family. The (in general unknown) ground state can have many metastable configurations nearby and the use of the majority-rule may trap the system in one of them.

As a way of quantifying the coherence of the global response to the forcing, we chose the spectral amplification factor $R$, defined as the ratio of the output to input power at the corresponding driving frequency\cite{ampli}:
\begin{equation}
R=\left\langle\frac{4}{a^{2}}\left|\left\langle\!\left\langle {\bf \tt m}(t)e^{-i\Omega t}\right\rangle\!\right\rangle \right|^{2}\right\rangle,
\end{equation}
where $\left\langle\!\left\langle ...\right\rangle\!\right\rangle$ is a time average, ${\bf \tt m(t)}$ is the global response (system's magnetization):
\begin{equation}
\label{defm}
{\bf \tt m}(t)=\frac{1}{N}\sum_{i=1}^{N}\mu_{i}(t),
\end{equation}
and $\left\langle ...\right\rangle$ is an ensemble average over network realizations, initial conditions and realizations of the dynamics. Large values for $R$ indicate that the global variable ${\bf \tt m}(t)$ follows the external forcing, while small values of $R$ indicate a small influence of the forcing on the global variable.

\subsection{Simulation results}
\label{sec:2.2}

\begin{figure}
\vspace{0.5cm}
\includegraphics[width=6cm]{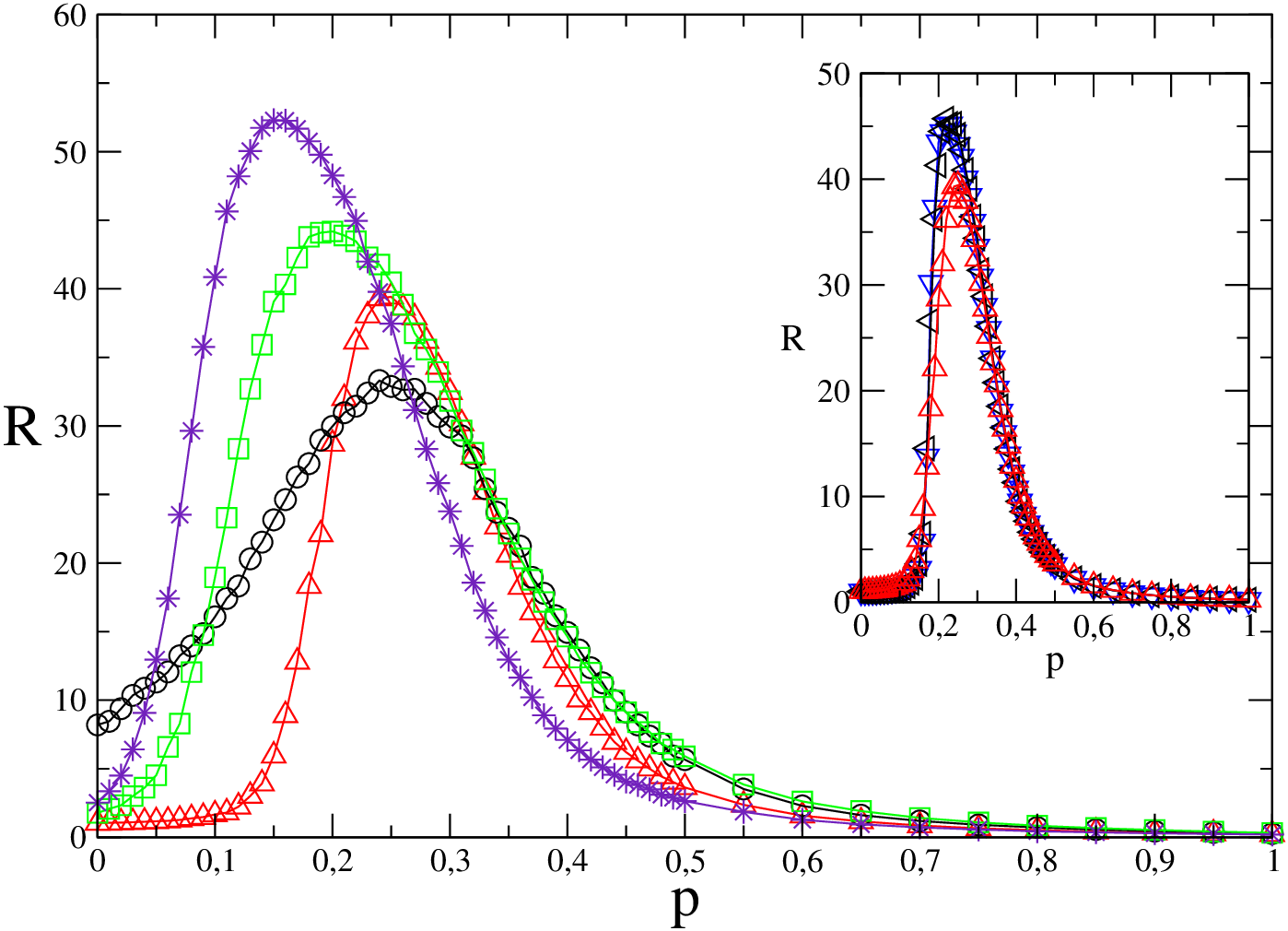}
\caption{Spectral amplification factor $R$ versus probability of repulsive links $p$. Parameters are: $a=0.15$, $\Omega=\frac{2\pi}{100}$, $\kappa=1$. In the main graph, $N=100$ and symbols correspond to topologies:  ring with $k=10$ neighbors ($\circ$),  square lattice  with $k=8$ neighbors in the Moore neighborhood ($\square$), and random networks with average number of neighbors $k=10$ and rewiring probability $q=0.2$ ($\ast$) and $q=1$  ($\vartriangle$). In the inset, we chose the random network with  $q=1$, $k=10$, and different curves correspond to sizes $N=100$ ($\vartriangle$), $500$ ($\triangleleft$), and $1000$ ($\triangledown$).}
\label{f1saf}
\end{figure}

\begin{figure}[h!]
\vspace{0.5cm}

\resizebox{0.40\textwidth}{!}{%
  \includegraphics{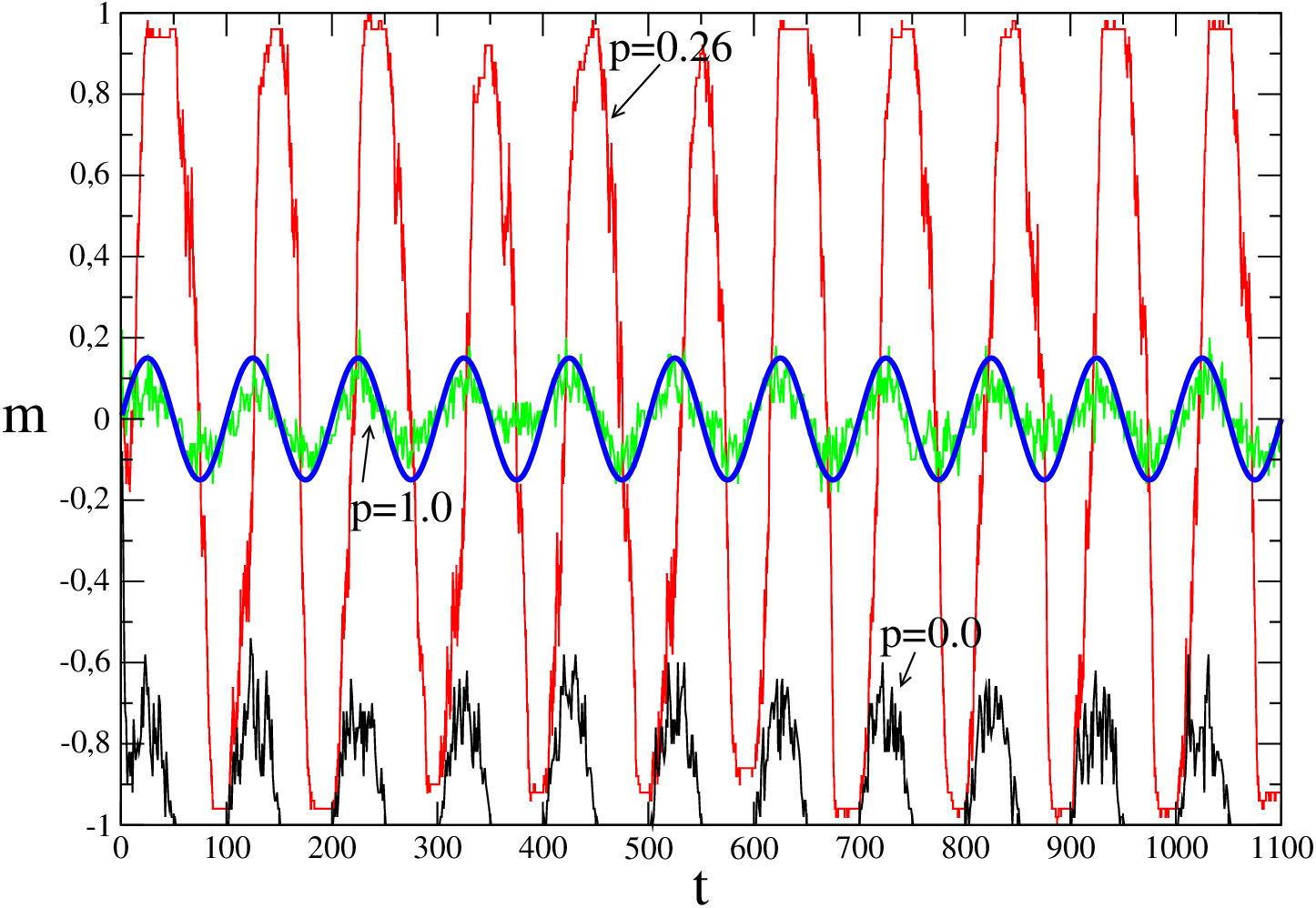}
}
\caption{Evolution of magnetization in time (random network, $q=1$, $k=10$). Other parameters are: $N=100$, $a=0.15$, $\Omega=\frac{2\pi}{100}$, $\kappa=1$.}
\label{f2serie}
\end{figure}

The main result of this paper is that there is a resonance effect, a maximum of the amplification factor $R$, at an intermediate value of the probability of repulsive links $p$, as shown in figure  \ref{f1saf}.  The existence of this maximum is also visible when looking at the amplitude of the oscillations of the global variable ${\bf \tt m}(t)$ -- figure \ref{f2serie}. For small $p$,  ${\bf \tt m}(t)$ oscillates with a small amplitude (of order $a$) around a value close to either $+1$ or $-1$. As $p$ increases, one clearly notices that the amplitude increases dramatically and ${\bf \tt m}(t)$ oscillates around $0$. As $p$ increases even further, the amplitude of the oscillations decreases but the global variable still oscillates around $0$. This resonance effect appears for all lattices considered, regular or random, for all values of the rewiring probability $q$.

\begin{figure}[h!]
\vspace{0.5cm}

\resizebox{0.40\textwidth}{!}{%
  \includegraphics{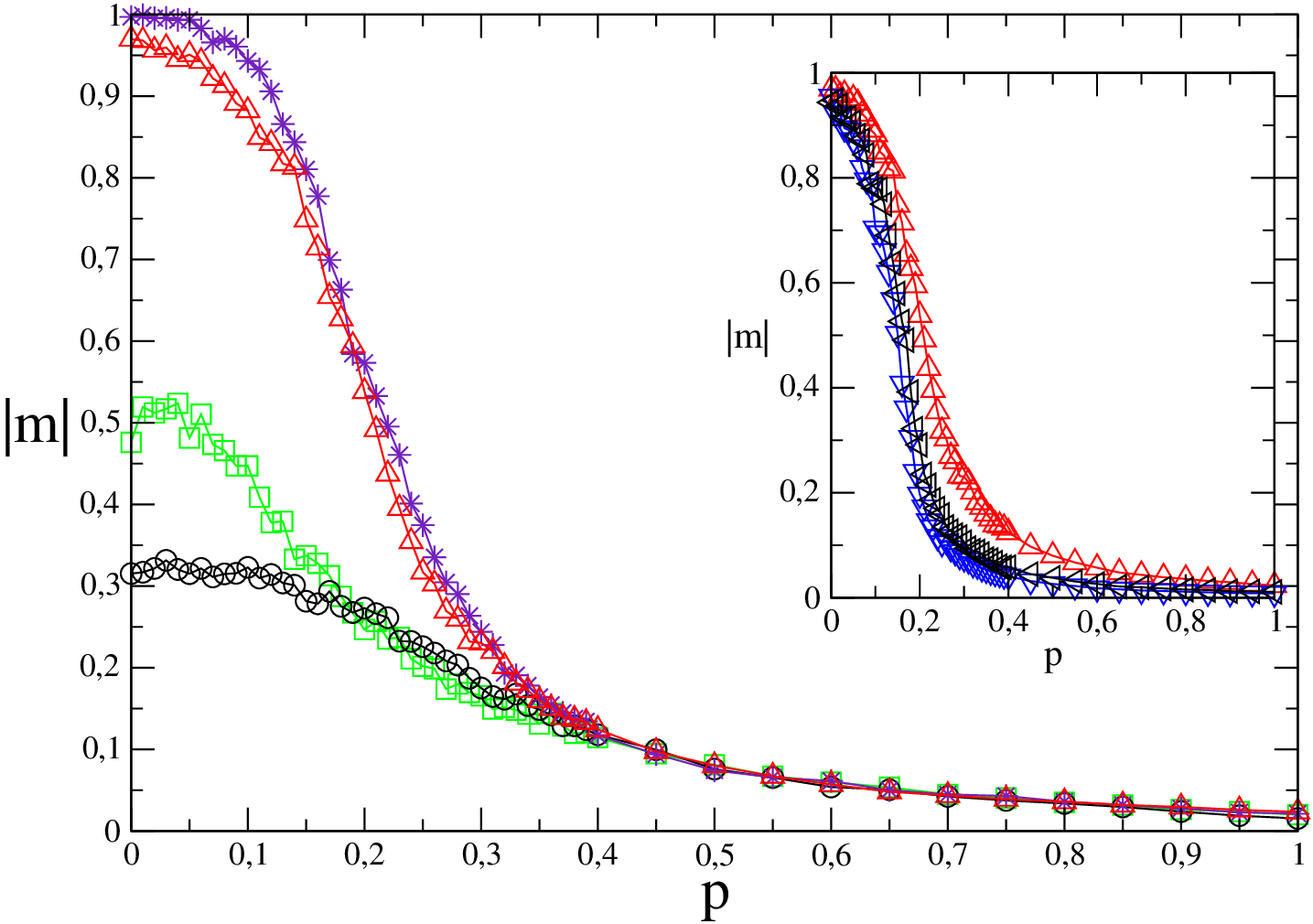}
}
\caption{Modulus of the average magnetization as a function of the probability of repulsive links. Same parameter values and symbol meanings than in  in figure \ref{f1saf}. In the regular networks, the existence of metaestable states reveals itself in a smaller magnetization at $p=0$.}
\label{f3mag} 
\end{figure}

As argued in \cite{TMTG06}, the existence of this resonance effect is the result of a degradation of order. In our case, the degradation of order has its origin in the increasing importance of the inhibitory connections. This is clearly seen in figure \ref{f3mag} where we plot the standard order parameter $m=\langle {\bf \tt m}(t)\rangle$ as a function of the probability $p$ of inhibitory links. The optimal probability for resonance $p_c$ (location of the peak of figure \ref{f1saf}) is found near the phase transition between the ferro and paramagnetic regions.

The existence of this order-disorder transition and its relation to the resonance effects are reproduced by a simple mean-field theory that we develop in some detail in the next section.

\section{Mean-field approach}
\label{sec:3}

At each time step the magnetization ${\bf \tt m}(t)$ may change due to the modification of a single variable $\mu_i$. The following relation holds exactly for the ensemble average $m(t)$:
\begin{equation}
Nm(t+\tau)=N m(t)+\left\langle \mu_i(t+\tau)-\mu_i(t)|\{\mu(t)\}\right\rangle
\end{equation}
where $\{\mu(t)\}=(\mu_1(t),\dots,\mu_N(t))$ denotes the particular realization of the $\mu_i$ variables and $\langle\dots|\dots\rangle$ denotes a conditional ensemble average. By identifying $\tau=1/N$ and rearranging we get:
\begin{eqnarray}
\frac{m(t+\tau)-m(t)}{\tau}&=&\left\langle \mu_i(t+\tau)-\mu_i(t)|\{\mu(t)\}\right\rangle=\nonumber\\
&&-m(t)+\left\langle \mu_i(t+\tau)|\{\mu(t)\}\right\rangle
\end{eqnarray}
We now identify the left hand side as the time derivative and use the dynamical rules given by Eq.(\ref{par}) to write:
\begin{eqnarray}
\frac{d m(t)}{dt}&&=-m(t)+|f(t)|\left\langle{\rm sign}[f(t)] |\{\mu(t)\}\right\rangle+\nonumber\\ &&(1-|f(t)|)\left\langle {\rm sign}\left[\sum_{j\in V(i)}\omega_{ij}\mu_{j}(t)\right] |\{\mu(t)\}\right\rangle
\end{eqnarray}
where we have used the notation $f(t)=a\sin(\Omega t)$. Since the forcing $f(t)$ is independent of the state $\{\mu\}$, then $\left\langle {\rm sign}[f(t)] |\{\mu(t)\}\right\rangle={\rm sign}[f(t)] $. Moreover $|f(t)| {\rm sign}[f(t)]=f(t)$. For the last term of the right hand side of this equation we use the mean-field approximation:
\begin{equation}
\sum_{j\in V(i)}\omega_{ij}\mu_{j}(t)\approx \left[\sum_{j\in V(i)}\omega_{ij}\right]\cdot m(t)
\end{equation}
where we replace the value $\mu_j(t)$ by the average value $m(t)$.

Now $\sum_{j\in V(i)}\omega_{ij}=k_i^+-\kappa k_i^-=k_i^+(1+\kappa)-k_i\kappa$, and the mean-field approximation can be rewritten as:
\begin{eqnarray}
&&\left\langle {\rm sign}\left[\sum_{j\in V(i)}\omega_{ij}\mu_{j}(t)\right] |\{\mu(t)\}\right\rangle=\nonumber\\ 
&& (-1)\cdot Prob\left(\left[k_i^+(1+\kappa)-k_i\kappa\right]m(t)<0\right)+\nonumber\\
&& (+1)\cdot Prob\left(\left[k_i^+(1+\kappa)-k_i\kappa\right]m(t)>0\right)\nonumber\\  
&& =1-2Prob\left(\left[k_i^+(1+\kappa)-k_i\kappa\right]m(t)<0\right)\nonumber\\ 
&& \equiv G(m(t))
\end{eqnarray}
from where we obtain the desired mean-field equation:
\begin{equation}
\label{mfem}
\frac{d m(t)}{dt}=-m(t)+f(t)+(1-|f(t)|)G(m(t))
\end{equation}
The function $G(m)$ can be easily computed in terms of the cumulative probability function $F_{k_i}$ of the binomial distribution of the number of positive links, given that the total number of links is $k_i$. This is precisely defined as:
\begin{equation}
F_k(x)=\sum_{k^+< x}{k\choose k^+}p^{k-k^+}(1-p)^{k^+}.
\end{equation}
In the case $m>0$, 
\begin{eqnarray}
&&Prob\left(\left[k_i^+(1+\kappa)-k_i\kappa\right]m(t)<0\right)=\nonumber\\ 
&&Prob\left(k_i^+< \frac{k_i\kappa}{1+\kappa}\right)=F_{k_i}\left(\frac{k_i\kappa}{1+\kappa}\right),
\end{eqnarray}
while, for $m<0$,
\begin{eqnarray}
&&Prob\left(\left[k_i^+(1+\kappa)-k_i\kappa\right]m(t)<0\right)=\nonumber\\ 
&&Prob\left(k_i^+> \frac{k_i\kappa}{1+\kappa}\right)=1-F_{k_i}\left(\frac{k_i\kappa}{1+\kappa}\right).
\end{eqnarray}
By averaging over the distribution of the number of neighbors, we get:
\begin{equation}
G(m)={\rm sign}(m)\sum_{k_i} P_{k_i} \left[1-2F_{k_i}\left(\frac{k_i\kappa}{1+\kappa}\right)\right]
\end{equation}
$P_{k_i}$ being the probability that a node has $k_i$ links. Within the spirit of the mean-field approximation we assume that all nodes have the same number of links $k_i=k$ and replace the above formula by:
\begin{equation}
\label{eq:gm}
G(m)={\rm sign}(m)\left[1-2F_{k}\left(\frac{k\kappa}{1+\kappa}\right)\right].
\end{equation}

\begin{figure}[h]
\vspace{0.5cm}

\resizebox{0.40\textwidth}{!}{%
 \includegraphics{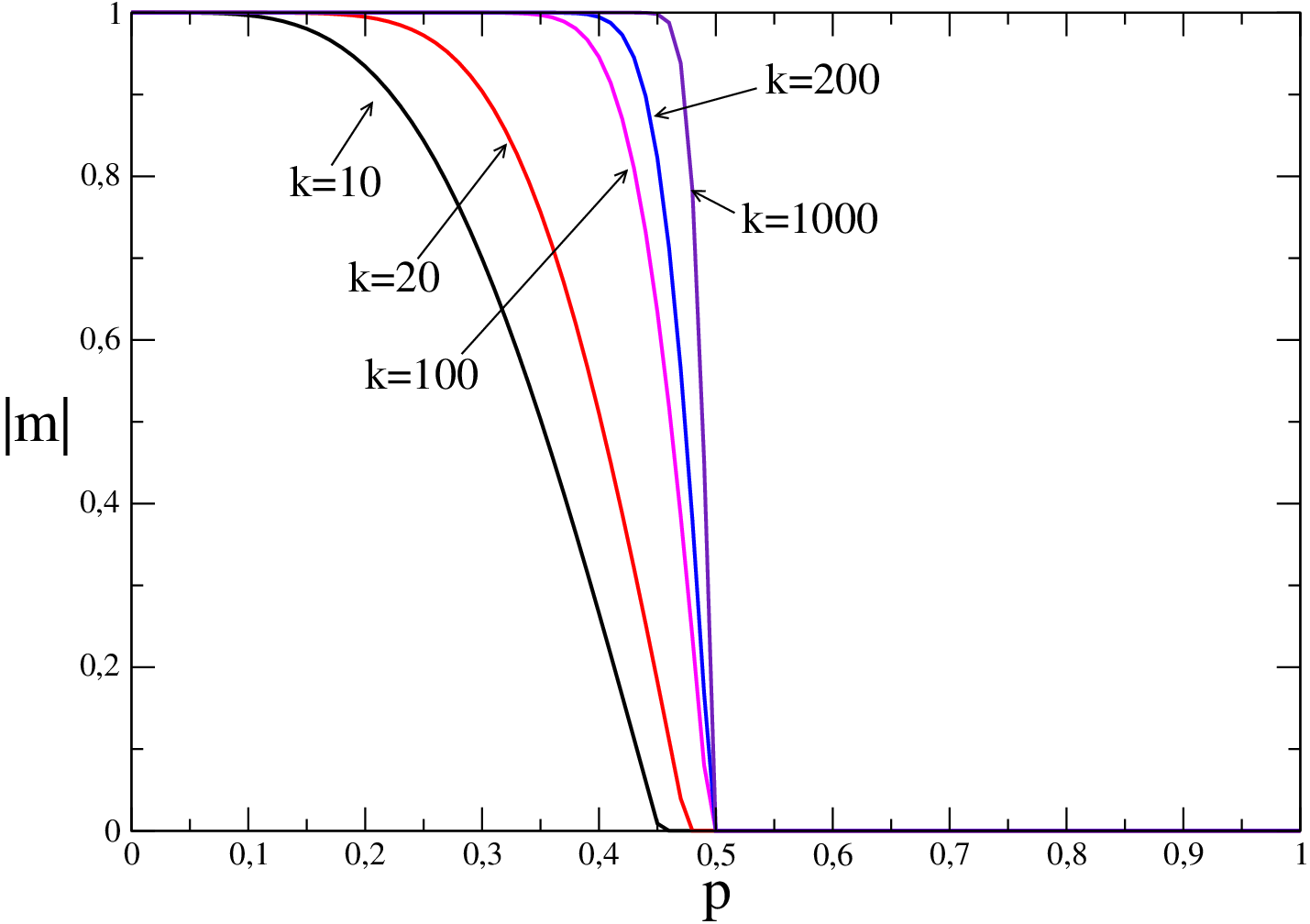}
}
\caption{Modulus of the average magnetization as a function of the probability of repulsive links according to the mean-field theory for $\kappa=1$.}
\label{f4mmfi}       
\end{figure}

In case of no forcing, $f(t)=0$, the equilibrium value $m_0$ of the magnetization satisfies $m_0=G(m_0)$. A standard analysis of this equation predicts a phase transition separating a regime of non-zero stable solutions  $\pm m_0 \neq 0$ from a regime in which the only solution is $m_0=0$. The coexistence line is $m_0=1-2F_{k}\left(\frac{k\kappa}{1+\kappa}\right)$ and the critical point occurs at $F_{k}\left(\frac{k\kappa}{1+\kappa}\right)=1/2$. In figure \ref{f4mmfi} we plot the equilibrium magnetization $m_0$ as a function of the probability $p$ for fixed $k$. It is clear from this figure that the mean-field approximation reproduces the loss of order that arises as the proportion $p$ of negative links increases, although the precise location of the transition point is not well reproduced. 

\begin{figure}[h]
\vspace{0.5cm}

\resizebox{0.40\textwidth}{!}{%
  \includegraphics{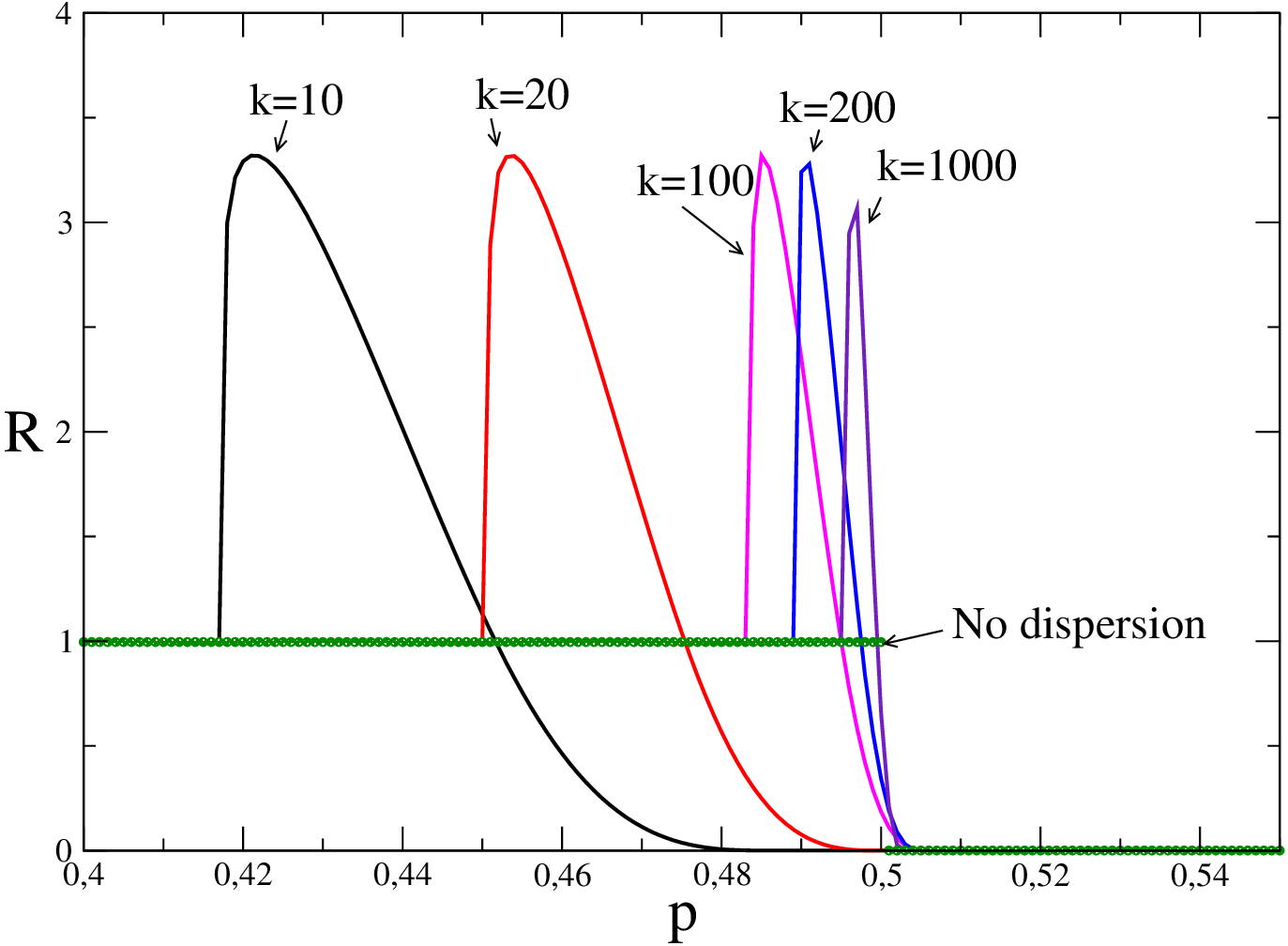}
}
\caption{Spectral amplification factor versus probability of repulsive links according to the mean-field theory for $a=0.15$, $\Omega=\frac{2\pi}{100}$, $\kappa=1$.}
\label{f5safmfi}       
\end{figure}

In figure \ref{f5safmfi} we plot the amplification factor computed after a numerical integration of Eq.(\ref{mfem}). Qualitatively, the results agree with those of simulations presented in the previous section: there is a resonance effect, i.e. the response shows a maximum as a function of $p$. The maximum value is reached for a value $p_c$, close to that signaling the order-disorder transition. Furthermore, it can be noticed that the size of the amplification region, defined as the set of values of $p$ for which $R>1$, is similar to the size of the transition region, defined roughly as the set of values of $p$ for which the magnetization satisfies $m(p)<0.5$ and the maximum is achieved at a value of $p$ such that $m(p)\approx 0.2-0.3$. As the average number of neighbors $k$ increases, the size $\Delta p$ of this region decreases as $k^{-1/2}$ and it disappears in the limit $k\to\infty$. Since the relative dispersion in the number of positive links also scales as $\sigma[k^+]/\langle k^+\rangle\sim  k^{-1/2}$, one is tempted to attribute the existence of the resonance to the existence of such a dispersion, a fact already stressed in the study of synchronized oscillations induced by diversity\cite{TZT08}. This is supported by a modified version of the mean-field approach in which the dispersion is strictly equal to $0$. This can be achieved by using in (\ref{eq:gm}) the probability distribution that would arise if all nodes had the same number $k_i^+$ of positive links, namely $F_{k}(x)=0$ if $x<pk$ and $F_{k}(x)=1$ if $x>pk$.  As shown in figure \ref{f5safmfi}, in this case the amplification region has disappeared altogether.

\section{Mechanism}
\label{sec:4}

\subsection{Microscopic point of view}
\label{micro}

We now give an explanation of some features of the observed resonance from a microscopic point of view, i.e. analyzing the evolution of individual values of $\mu_i$.

According to the rules (\ref{par}), a chosen node takes the sign of the external signal with a probability $|a\sin(\Omega t)|$, independently of the current system configuration. To enhance resonance, there are two necessary requirements  after a node has changed its state: to maintain the perturbation in the next time steps, and to spread it to its neighbors. The crucial issue is then how the local configuration of nodes and links helps (or hinders) this ordering process.

To spread a perturbation, it would be an advantage to have all-attractive couplings; however, to maintain its state, the node cannot be too constrained by its neighbors. With a high homogeneity of the neighbors states and a positive connection with all of them, a perturbed spin would likely be forced to go back to its original state next time it is selected. At the other extreme, when all its connections are negative, a perturbed node is also very much constrained by the state of its neighbors, the local field being maximal for a local anti-ferromagnetic ordering.
At an intermediate level of positive and repulsive connections, we have the optimal state. It has a capacity to spread a perturbation to the whole network, but constrains minimally a node that has been perturbed. Due to the combination of attractive and repulsive links, the local field around a node is close to zero. Therefore, if a node changes its state, it possibly won't be forced to return to  its previous position after consulting with its neighbors. On the other hand, it is easy to spread a perturbation: if a node had previously a zero local field, after one neighbor has changed, the balance is broken, and it has to align with that neighbor, if the connection is positive. 

\begin{figure}[h]
\vspace{0.5cm}

\resizebox{0.40\textwidth}{!}{%
  \includegraphics{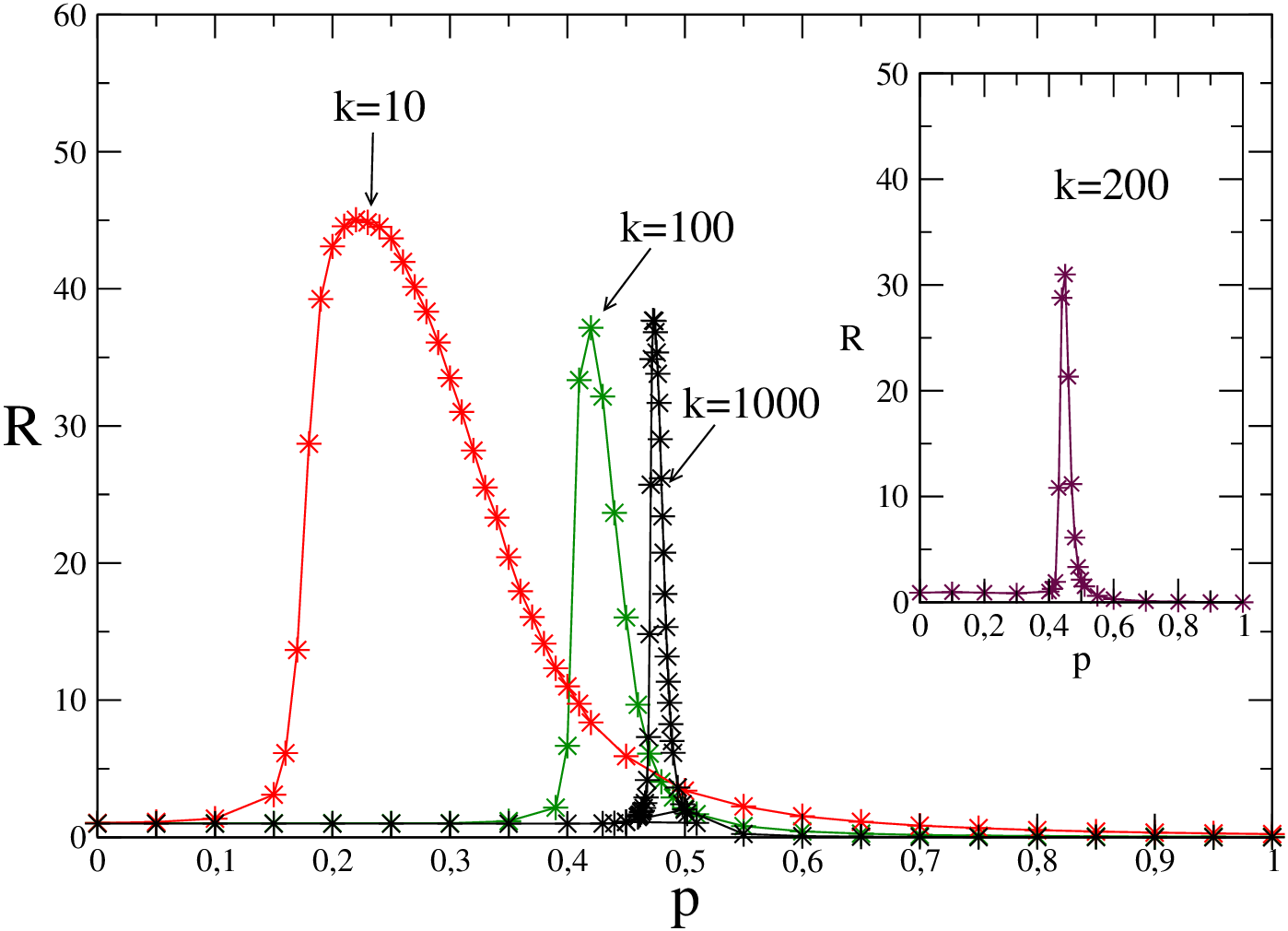} 
}
\caption{Spectral amplification factor versus probability of repulsive links for a random network with $q=1$,  $a=0.15$, $\Omega=\frac{2\pi}{100}$, $\kappa=1$. Main graph uses $N=1001$ while the inset shows the case $N=201$.}
\label{f6safn}       
\end{figure}

This microscopic picture will help us to understand some of the observed features. For example, in figure ~\ref{f6safn} we show that the amplification region $\Delta p$ decreases with increasing $k$ whereas $p_c$ tends to $0.5$. Both facts agree qualitatively with the predictions of the mean-field theory. It is clear that for large $k$ the condition of a local field close to zero can only be satisfied for a probability of repulsive links near $0.5$. This is easily illustrated when one considers the case of $p$ far from $0.5$ and a uniform magnetization (at the peak of a signal's cycle). Getting a local field close to zero when the connectivity is high requires many neighbors flips. Since the unit to be updated is chosen randomly at each time step, it is likely that a unit is chosen twice before enough of its neighbors have been perturbed. On the other hand, $p=0.5$ is the upper limit for the amplification region, because a majority of positive links is necessary to have perturbation spreading. As the proportion of repulsive links approaches $0.5$, more neighbors have a negative connection  and they will exert, when perturbed, an influence opposite to the signal. 

Note that for the resonance to disappear we need formally the limit $k \to\infty$. In a finite network, the maximum value is $k=N-1$ and, as shown in figure  ~\ref{f6safn} for $N=201$ and $N=1001$, the resonance does not disappear completely even for the maximum connectivity.

\begin{figure}[h]
\vspace{0.5cm}

\resizebox{0.40\textwidth}{!}{%
  \includegraphics{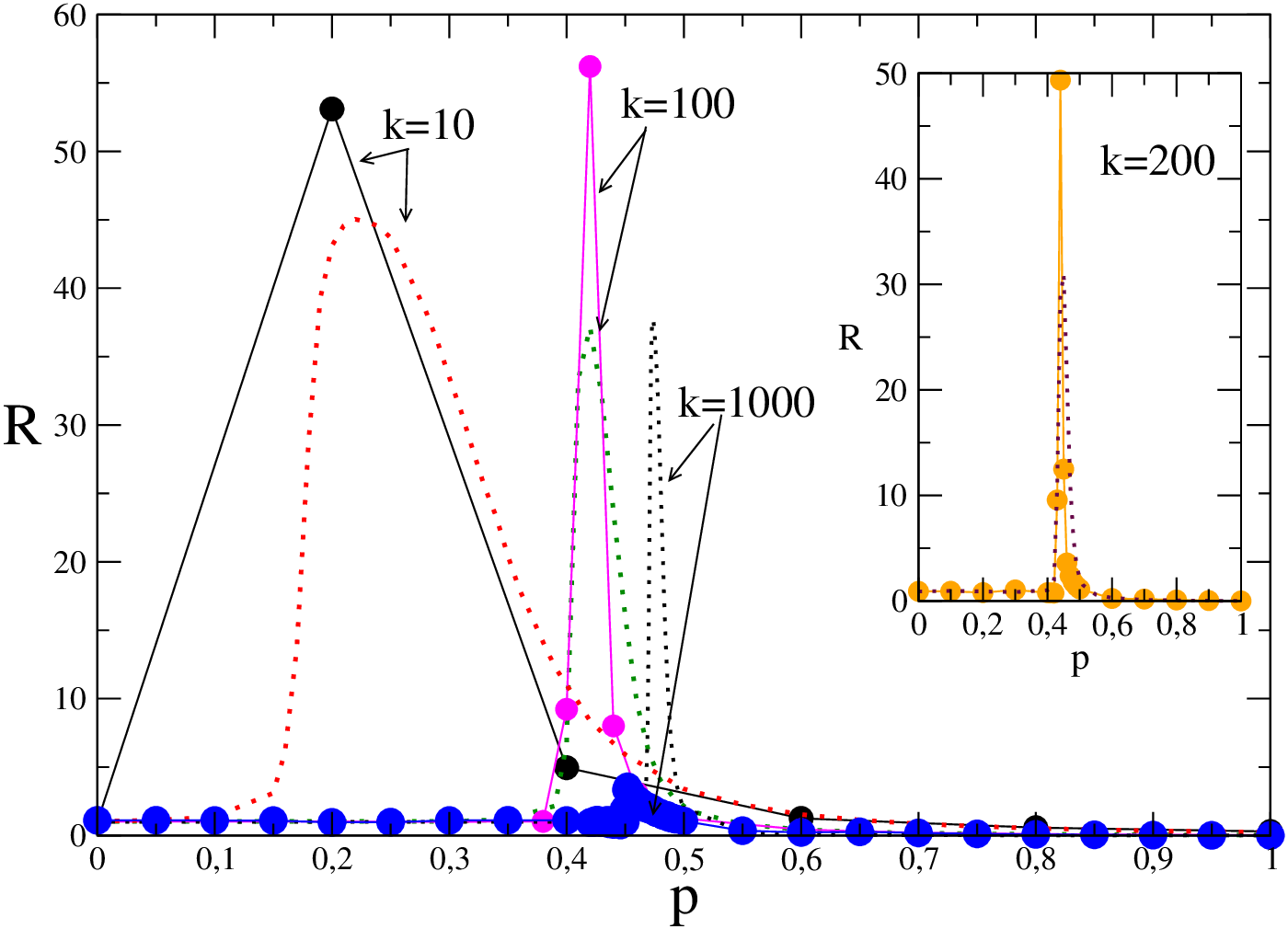} 
}
\caption{Spectral amplification factor versus probability of repulsive links for a ``no-dispersion" network in which all sites have the same number of positive and negative links. Due to the particular way the network is constructed\cite{MS}, only values of $p=k/N$ where the total number of neighgbors per site, $k$, is an even integer number are allowed. Parameters are $N=1001$, $a=0.15$, $\Omega=\frac{2\pi}{100}$, $\kappa=1$ (main graph) and $N=201$ (inset). Note that the amplification region shrinks as $k$ increases.  For comparison, we also include as dotted lines the results of figure \ref{f6safn}.}
\label{f7safnc}      
\end{figure}

As we did in the mean-field treatment, and in order to isolate the influence of competitive interactions from the disorder induced by the dispersion in the number of links, we also present in figure  ~\ref{f7safnc} results from random networks when all nodes have exactly the same number of neighbors $k$ and the same proportion \emph{p} of repulsive links \cite{MS}. At variance with the previous results, an almost total reduction of the amplification region can be achieved even for finite values of $N$, for large enough $k$. This proves that diversity in the number of positive links is an important ingredient for the resonance effect.

Why does dispersion matter? The precise mechanism is hard to grasp, but it is certainly related to a degradation of order at local level. To decrease the chance of having perturbed neighbors driving several units in the direction opposite to the signal, there have to be many nodes with a clear majority of positive links. But as we saw above -- assuming every node had the same number of negative links -- those units require many neighbors flips, to maintain their local field close to zero. However, if the nodes are heterogeneous, an unit with a lower than average number of repulsive links can profit from those neighbors that have many negative connections to other nodes. Since those are more susceptible to changes, their presence decreases the local field, thereby diminishing the need for many neighbors updates.
This result confirms the importance of diversity in making the phenomenon more robust, but also shows that we can have an amplification even without diversity.

\subsection{Macroscopic point of view}
\label{macro}

In this subsection, we consider the explanation of the resonance from the macroscopic point of view, i.e. we look at the behavior of the collective variable (magnetization) $m(t)$. We assume that the dynamics of this macroscopic variable in the no-forcing case, $f=0$, can be described in terms of relaxation in a potential function $V(m)$. The absolute minima $\pm m_0$ of the potential give the rest states which are separated by a potential barrier $\Delta V$. This picture has proved to be valid in other problems with diversity in the parameters \cite{TMTG06} and it certainly holds in the mean-field limit where, according to the previous section, the dynamical equation is $\frac{dm}{dt}=-\frac{dV}{dm}$ with a potential $V(m)=\frac{m^2}{2}-M(p) |m|$ with  $M(p)=1-2F_{k}\left(\frac{k\kappa}{1+\kappa}\right)$ running from $M(0)=1$ to $M(1)=-1$. There are two minima of the potential, $m_0=\pm M(p)$ for $M(p)>0$, and a single minimum $m_0=0$ for $M(p)<0$, or $p>p_c$, the critical point. For small $p$ the barrier separating the two minima $\Delta V=\frac{M(p)^2}{2}$ is high and it can not be overcome by the effect of the weak forcing $f(t)$. The only effect of the forcing is a small oscillation around one of the minima (chosen by the initial conditions). As $p$ increases, the two minima of the potential get closer to each other and the barrier separating them decreases such that, at a particular value of $p$ the forcing is able to overcome the barrier and $m(t)$ oscillates between the two minima $\pm m_0$. As $p$ crosses the critical value $p_c$, the two minima merge at $m_0=0$, the barrier disappears and the effect of the forcing is reduced again to small oscillations around a single minimum.

\begin{figure}[h]
\vspace{0.5cm}

\resizebox{0.40\textwidth}{!}{%
\includegraphics{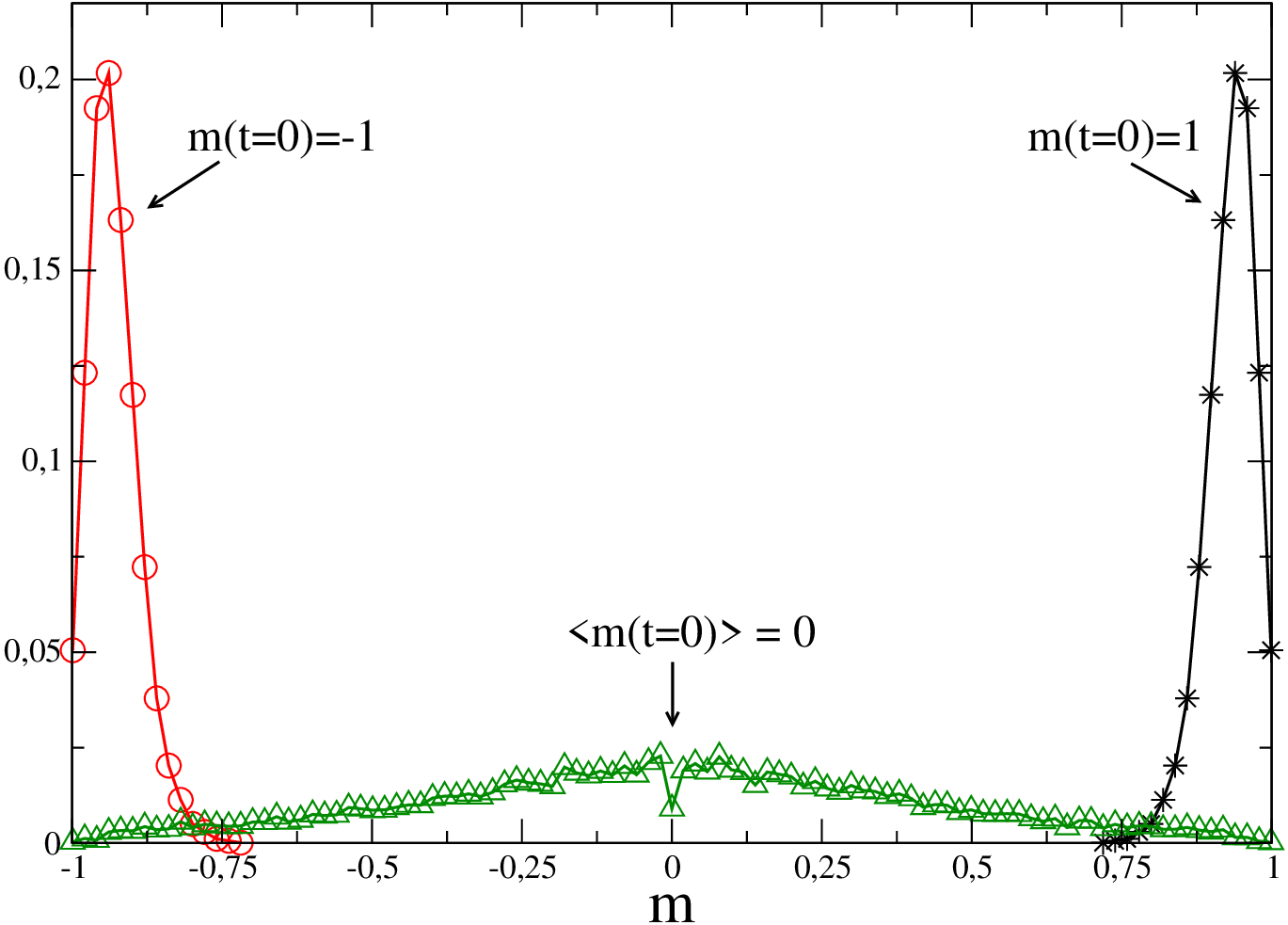}
}
\caption{Distribution of stable states at the optimal probability $p_c=0.25$ in the case of an unforced random network with $q=1$, $N=100$, $k=10$, $\kappa=1$ starting from three different inital conditions: all spins equal to $+1$ (data set indicated as $m(t=0)=1$), all spins equal to $-1$ ($m(t=0)=-1$) and spins take randomly the value $\pm 1$ ($\langle m(t=0)\rangle=0$).}
\label{f8qdistri}       
\end{figure}

\begin{figure}[h]
\vspace{0.5cm}

\resizebox{0.40\textwidth}{!}{%
  \includegraphics{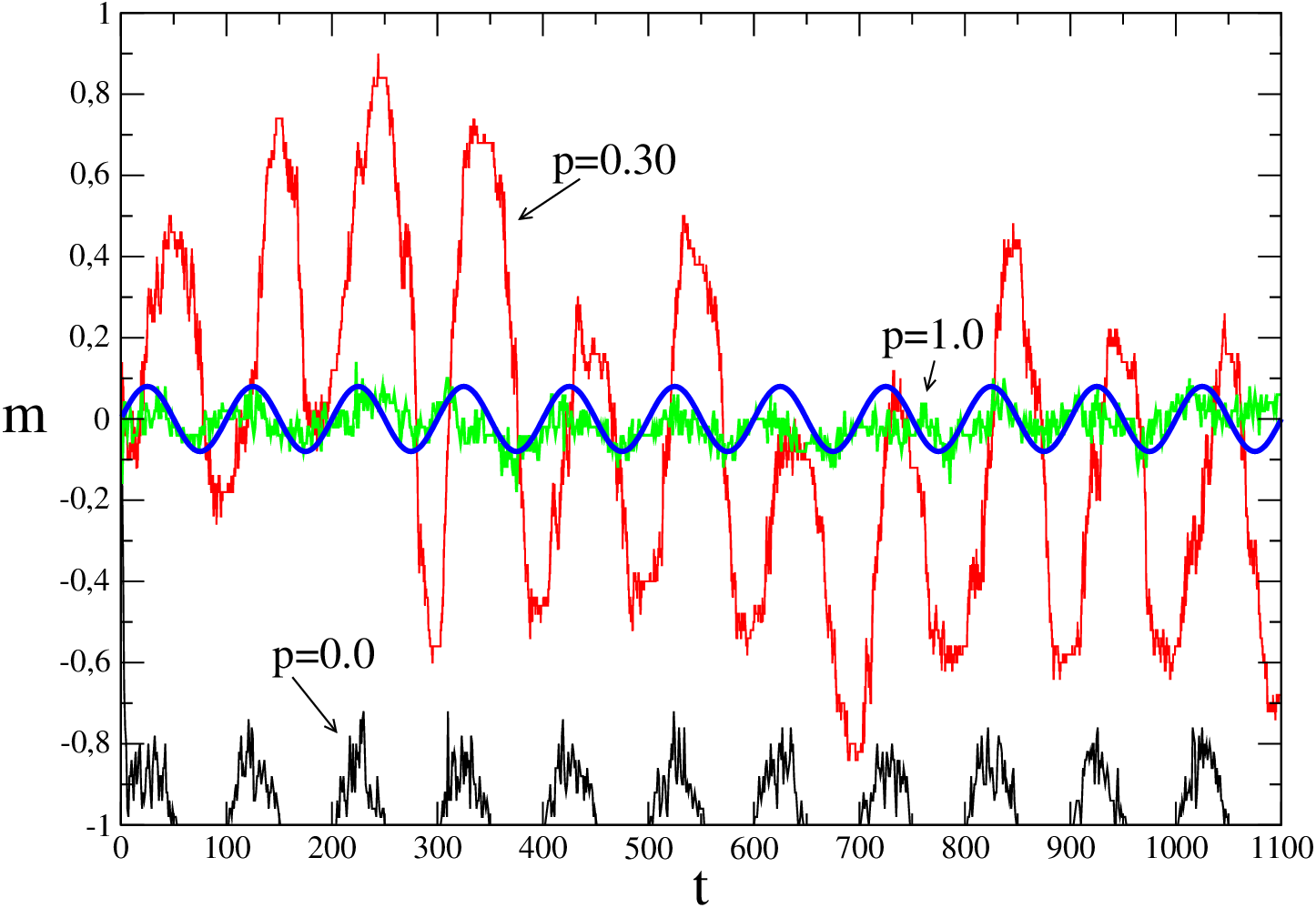}
}
\caption{Evolution of the magnetization following a weak signal $a=0.08$, $\Omega=\frac{2\pi}{100}$, in the case of a random network, $q=1$,  $k=10$,  $\kappa=1$.}
\label{f9008p100}       
\end{figure}

\begin{figure}[h]
 \vspace{0.5cm}

\resizebox{0.40\textwidth}{!}{%
  \includegraphics{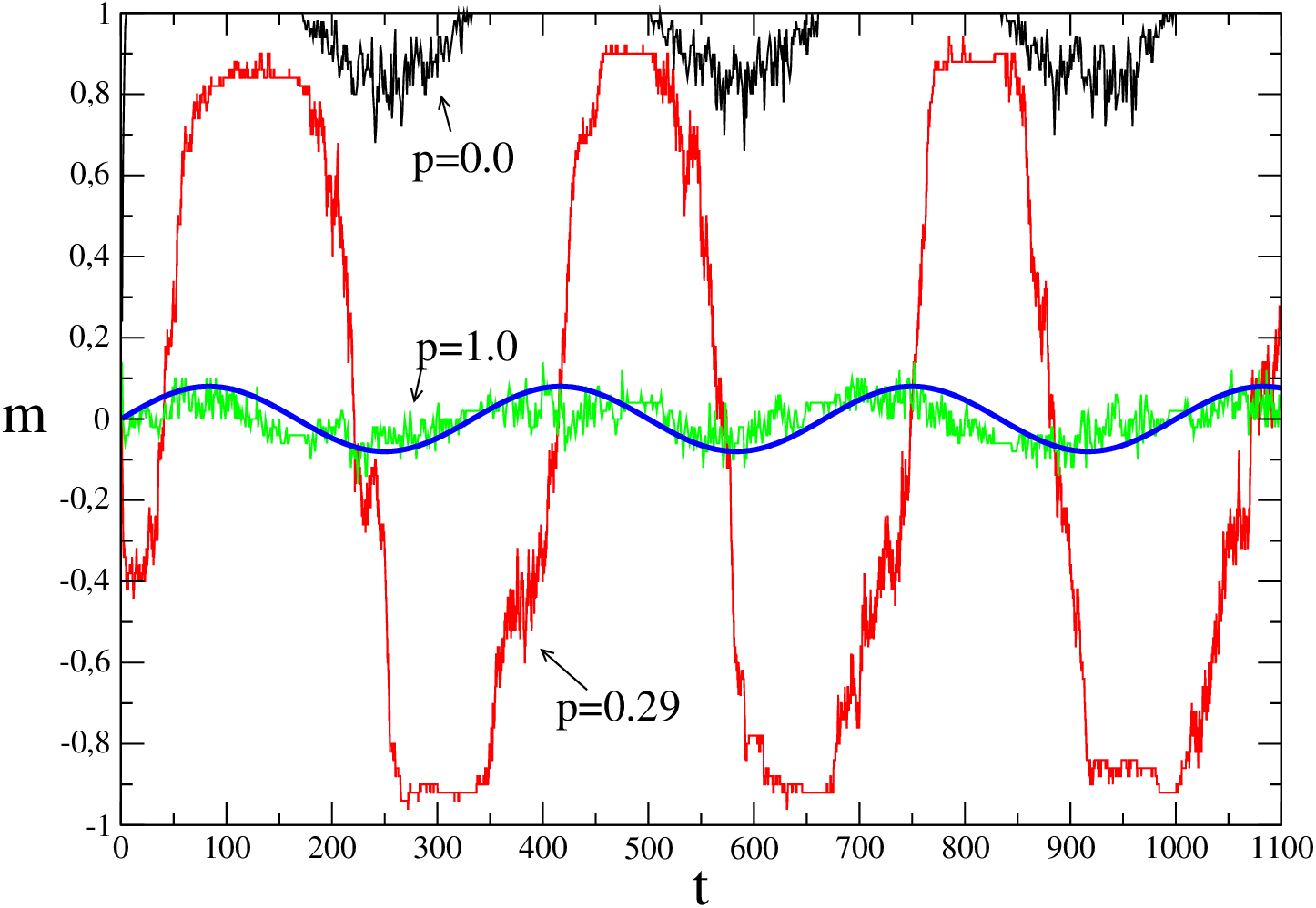}
}
\caption{Same as figure \ref{f9008p100} for a slower forcing $\Omega=\frac{2\pi}{333.3}$.}
\label{f10008p333}      
\end{figure}

To apply this potential image beyond the mean-field approximation we need to include an important modification. As discussed before, the energy landscape is that of a spin-glass with many metastable states and two absolute minima $\pm m_0$. As a consequence, in the no-forcing case, the final state reached depends strongly on initial conditions. This is illustrated in figure~\ref{f8qdistri} where we plot the probability distribution of the final magnetization. If the initial state is the ordered state $\mu_i=+1$ (resp. $-1$) $\forall i$, the final magnetization is peaked near $m=+1$ (resp. $-1$). If the initial state adopts $\mu_i=\pm1$ randomly, then the final magnetization is peaked around $m=0$. This reflects the existence of many barriers separating the metastable states from the absolute minima of the potential. When the forcing is introduced, it has to be able to overcome all these intermediate barriers. The final image is that of a particle moving in a ``rugged" potential. As $p$ increases, the height of those barriers decreases and the forcing is able to explore a larger fraction of the configuration space, but not necessarily leading to trajectories  ending in the absolute minima of the potential. This can be seen in figure~\ref{f9008p100} where we show the effect of a forcing weaker than that used in figure \ref{f2serie}. The magnetization oscillates around a mean value that drifts with time. If we enlarge the period of the forcing -- figure~\ref{f10008p333} -- the oscillations become wider and the system has now enough time to reach the equilibrium minima close to $m_0=\pm 1$. \cite{transient}

\section{Conclusion}
\label{sec:5}
We have used Monte Carlo simulations and analytical (mean field) calculations to investigate the response of a system of two-state units, with both attractive and repulsive interactions and majority-rule dynamics, to a weak periodic signal. For both regular and random networks, we have found that competing interactions can enhance the system response -- a kind of ``divide and conquer'' strategy. In each case, a resonance was found for an optimal percentage of negative links which depends on the model parameters. Applications include opinion dynamics and neuron networks but the model is generic enough to predict that the same type of effect can be found in other systems.

We have discussed in some detail the microscopic mechanism for the amplification. We argued that the flexibility of the system to follow the external signal requires that the {\it local field} seen by each unit is kept close to zero and analyzed how this condition might be achieved in some parameter limits.

A macroscopic analysis, in terms of a relaxation dynamics in a bistable potential, is able to explain the mean-field results. It is difficult
to use this description beyond the mean field treatment, due to the presence of many metastable configurations.  Because of their presence,
a large response, corresponding to oscillations around (symmetrical) absolute minima can be obtained for a sufficiently slow forcing. There are studies that point to the role network topology plays in synchronization or response to stimuli \cite{BE}. Analyzing the effect of coupling strength, degree distribution and other network characteristics on the coherent response may shed some light on how the mechanism can be optimized.

\bigskip
{\textbf{Acknowledgments:}}
We acknowledge financial support from project FIS2007-60327 from MEC (Spain).  T.V.M. acknowledges the support of FCT (Portugal) through Grant No.~SFRH/BD/23709/2005. 

%
% BibTeX users please use
% \bibliographystyle{}
% \bibliography{}

\begin{thebibliography}{}

\bibitem{benzi} R. Benzi, A. Sutera, and A. Vulpiani, J. Phys. A \textbf{14}, 453 (1981)

\bibitem{nicolis} C. Nicolis and G. Nicolis, Tellus \textbf{33}, 225  (1981)

\bibitem{lasers} B. McNamara, K. Wiesenfeld, and R. Roy, Phys. Rev. Lett. \textbf{60}, 2626 (1988)

\bibitem{SQUIDS} A. D. Hibbs, A. L. Singsaas, E. W. Jacobs, A. R. Bulsara and Bekkedahl J. J., J. Appl. Phys., \textbf{77} 2582 (1995); D. Gourier and D. Gerbault, Phys. Rev. B, \textbf{57},  2679 (1998).

\bibitem{neurons} J.K. Douglass, L. Wilkens, E. Pantazelou and F. Moss, Nature  \textbf{365}, 337 (1993) 

\bibitem{GHJM98} L. Gammaitoni, P. H\"anggi, P. Jung and F. Marchesoni. Rev. Mod. Phys. \textbf{70}, 223 (1998)

\bibitem{JSTAT} A. Bulsara, P. H\"anggi, F. Marchesoni, F. Moss and M. Shlesinger, eds, \textit{Stochastic Resonance in Physics and Biology}, J. Stat. Phys.  \textbf{70}, 1-512 (1993)

\bibitem{kramers} H. A. Kramers, \textit{Physica} \textbf{7},  284 (1940)

\bibitem{TMTG06} C. Tessone, C.R. Mirasso, R. Toral, J.D. Gunton, Phys. Rev. Lett. \textbf{97}, 194101, (2006)

\bibitem{TTV:2007} R. Toral, C. J. Tessone and J. Viana Lopes, Eur. Phys. J. Special Topics, \textbf{143}, 59 (2007)

\bibitem{array} J. F. Lindner, B. K. Meadows, W. L. Ditto, M. E. Inchiosa and A. Bulsara, Phys. Rev. Lett. \textbf{75}, 3 (1995)

\bibitem{Wio} H.S. Wio, Phys. Rev. E \textbf{54}, R3075 (1996)

\bibitem{ALA:2007} J.A. Acebr{\'on}, S. Lozano, A.A. Arenas, Phys. Rev. Lett. \textbf{99}, 128701 (2007).

\bibitem{CZL:2007} H. Cheng, J. Zhang, J. Liu, Phys. Rev. E \textbf{75}, 041910 (2007).

\bibitem{GGK:2007} M. Gassel, E. Glatt, F. Kaiser, Phys. Rev. E \textbf{76}, 016203 (2007).

\bibitem{Levya} I. Leyva, I. Sendi{\~n}a-Nadal, J. A. Almendral, and M. A. Sanjuán,  Phys. Rev. E, \textbf{74}, 056112 (2006)

\bibitem{TZT08} C.J. Tessone, D.H. Zanette, R. Toral,  Eur. Phys. J. B \textbf{62}, 319-326 (2008). 

\bibitem{shinomoto}Y. Shinomoto and Y. Kuramoto, Prog. Theor. Phys. \textbf{75}, 1105 (1986) 

\bibitem{julyan} J. Cartwright, Phys. Rev. E \textbf{62}, 1149 (2000)

\bibitem{GGK:2008} E. Glatt, M. Gassel, F. Kaiser. Europhys. Lett. \textbf{81}, 40004 (2008).

\bibitem{TSTC07} C.J. Tessone, A. Scire, R. Toral, P. Colet, Phys. Rev. E \textbf{75}, 016203 (2007)

\bibitem{KZ02} M. Kuperman, D. Zanette,  Eur. Phys. J. B, \textbf{26}, 387, (2002) 

\bibitem{watts} D. J. Watts and S. H. Strogatz, Nature {\bf 393}, 440 (1998)

\bibitem{LB:2000} D.P. Landau, K. Binder, {\sl A Guide to Monte Carlo Simulations in Statistical Physics}, Cambridge University Press (2000).

\bibitem{ampli} P. Jung and P. H\"anggi, Europhys. Lett. \textbf{8}, 505  (1989)

\bibitem{MS} We have adapted the ``local rewiring algorithm" (S. Maslov, K. Sneppen, Science \textbf{296}, 910 (2002)) to construct a random network where every node has exactly the same number of links $k$ and the same proportion $p$ of repulsive links -- the ``no dispersion" network. 

\bibitem{transient} We have also tried a version where we allow for a transient time: when a node is selected the connection to each neighbor is redefined as attractive or repulsive according to the given probability. After a large number of updates, we freeze the connections. We get a bimodal distribution suggestive of a bistable potential. 
In such case, perturbations induce a jump to the symmetric configuration. The reaction is stronger, but the drawback is a degraded periodicity in case of a very weak signal.  

\bibitem{BE} Y. Bar-Yam and I. R. Epstein, PNAS,  \textbf{101}, 4341 (2004) 


\end{thebibliography}

\end{document}